\documentclass[12pt]{article}
\usepackage{fullpage,epsfig,fancyheadings}
\usepackage[psamsfonts]{euscript}

\usepackage{epic}

\usepackage{amstex}

\newcommand{\beq}{\begin{equation}}
\newcommand{\eeq}{\end{equation}}

\begin{document}
\vspace*{-.6in}
\thispagestyle{empty}
\begin{flushright}
CALT-68-2049
DOE RESEARCH AND\\
DEVELOPMENT REPORT
\end{flushright}
\baselineskip = 20pt

\vspace{.5in}
{\Large
\begin{center}
The Decay $D^0\rightarrow \bar K^{*0} \pi^- e^+ \nu_e$ in the Context of Chiral Perturbation Theory\footnote{Work supported in part by the U.S. Dept. of Energy under Grant No. DE-FG03-92-ER40701.}
\end{center}}
\vspace{.4in}

\begin{center}
Hooman Davoudiasl\\
\emph{California Institute of Technology, Pasadena, CA  91125 USA}
\end{center}
\vspace{1in}

\begin{center}
\textbf{Abstract}
\end{center}
\begin{quotation}
\noindent We study the decay $D^0\rightarrow \bar K^{*0} \pi^- e^+ \nu_e$, 
using $SU(2)_L \otimes SU(2)_R$ chiral perturbation theory for heavy charmed mesons and vector mesons, in the kinematic regime where $p_M \cdot p_\pi/m_M$ (here $M = D^0$ or $\bar K^{*0}$) is much smaller than the chiral symmetry breaking scale, $\Lambda_{\chi SB}$ ( $\Lambda_{\chi SB} \sim $ 1 GeV).  We present the leading diagrams and amplitude, and calculate the rate, 
in the region where, to leading order in our calculations, 
the $\bar K^{*0}$ is at zero recoil in the $D^0$ rest 
frame.  The rate thus calculated is given in terms of a known form factor and depends on the $DD^* \pi$ coupling constant $g_D$ of the heavy (charmed) meson chiral perturbation theory Lagrangian.  A measurement of the above decay, in the aforementioned kinematic regime, can result in the extraction of an experimental value for $g_D$, accurate at the level of our approximations, and give us a measure of the validity of approaches based on chiral perturbation theory in studying similar processes.

\end{quotation}
\vfil

\newpage

\pagenumbering{arabic} 

\section{Introduction}

A sysytematic expansion for calculating a wide range of processes involving hadrons at low momenta is provided by chiral perturbation theory.  Apart from its uses in describing the strong interactions of the pseudo-Goldstone bosons (PGB's), that is, the octet of the lightest pseudo-scalar hadrons, chiral perturbation theory has been applied in describing the interactions of PGB's  with heavy matter fields, such as nucleons and heavy mesons containing a charm or a bottom quark \cite{tau1, tau2}.

Chiral perturbation theory, based on $SU(3)_L \otimes SU(3)_R$ chiral symmetry,  has also been applied to the interactions of PGB's with the lowest lying vector mesons, $\rho$'s, $K^*$'s, $\omega$, and $\phi$ \cite{Jenkins}.  
Ref. \cite{D.W.} makes use of this approach in studying the decay of the $\tau$ in the channels $\tau \rightarrow \rho \pi \nu_\tau$, $\tau \rightarrow K^*\pi\nu_\tau$, and $\tau \rightarrow \omega\pi\nu_\tau$.  This last decay 
mode is used to extract the absolute value of the $K^* K^* \pi$ coupling constant $\left|g^{(K^*)}_2\right|$, from experimental data \cite {CLEO}.  The decay rates in the above modes are calculated, using $SU(2)_L \otimes SU(2)_R$ chiral symmetry
only, thus not treating the strange quark mass as small.

In the present work, we use chiral perturbation 
theory for heavy charmed mesons \cite{Rapid} and heavy vector mesons  \cite{Jenkins}, in the limit of $SU(2)_L \otimes SU(2)_R$ chiral symmetry, 
to calculate the differential decay rate for $D^0\rightarrow \bar K^{*0} \pi^- e^+ \nu_e$ in a region of phase space where the pion is ``soft'' in the rest frames of both the $D^0$ and the $K^{*0}$.  Currently, there is an upper bound of 1.3\% on the branching ratio for this decay mode.  To calculate the leading order amplitude, we need the matrix element of the left-handed current $\bar s \gamma_\mu(1-\gamma_5)c$, between a $D^*$ and a $K^*$.  Since the form factors for this current are not measured, 
we relate them to the form factors of the same left-handed current between a 
$D$ and a $K^*$, using heavy quark symmetry \cite{Neubert, Zoltan}.  At the present time, one of the form factors of the $D\to K^*$ current is still unmeasured.  The dependence of the decay amplitude on this form factor is eliminated when we study the decay in the region where the $K^*$ is near zero recoil.  In this restricted region, the calculated differential decay rate depends on the $DD^*\pi$ coupling constant $g_D$ of heavy charmed meson chiral perturbation Lagrangian (coupling constant $g$ of Eq. (12) in 
Ref. \cite {Rapid}).   There is an experimental bound on the value 
of $g_D$, but the value of this constant remains to be measured.  There have been theoretical attempts at the determination of $g_D$, such as those involving radiative $D^*$ decays \cite{P, JF}.  In this paper, we present an indepent theoretical approach for obtaining the value of $g_D$.  The 
results of our paper can be compared to the data on  $D^0\rightarrow 
\bar K^{*0} \pi^- e^+ \nu_e$ to extract the experimental value of $g_D$ which can in turn provide a measure of the validity of 
the methods used here, by calculating other processes that depend on $g_D$ and
comparing with data.

We will introduce the chiral perturbation theory relevant to this work, in the next section.  In Section III, we present expressions for the left-handed hadronic currents for $D \to K^*$ and $D^* \to K^*$, and the leading amplitude for the decay.  Section IV contains the prediction for the differential decay rate, in a restricted region of phase space that will be discussed.  The concluding remarks are presented in Section V, followed by an appendix in which
some useful formulas are presented.

\section{Chiral Perturbation Theory}

In this section, we introduce the formalism necessary for the calculations discussed in this paper.  In the rest of this work, the words ``heavy meson'' refer to a meson containig a charm or a bottom quark, unless otherwise specified.  We will use notation similar to that of Refs. \cite {Rapid, D.W.}.  We start with the strong interactions of pions and heavy 
mesons, under $SU(2)_L \otimes SU(2)_R$ chiral symmetry.  The pions are incorporated into a $2\times2$ special
unitary matrix
\begin{equation}
\Sigma = \exp (2i \Pi/f_\pi),
\label{Sig}
\end{equation}
where
\begin{equation}
\Pi = \left[ \begin{array}{cc}
\pi^0/\sqrt{2} & \pi^+\\
\pi^- & - \pi^0/\sqrt{2}
\end{array} \right] .
\label{Pi}
\end{equation}

Under chiral $SU(2)_L \otimes SU(2)_R, \Sigma \rightarrow L\Sigma R^\dagger$,
where $L\in SU(2)_L$ and $R\in SU(2)_R$.  At the leading order in chiral perturbation theory, $f_\pi$ is given by the pion decay constant $f_\pi
\approx 132$ MeV.  To describe the interactions of pions with other fields, it is convenient to define
\begin{equation}
\xi \equiv \exp \left({i\Pi\over f_\pi}\right) = \sqrt{\Sigma}.
\label{xi}
\end{equation}
Under chiral $SU(2)_L \otimes SU(2)_R$,
\begin{equation}
\xi \rightarrow L \xi U^\dagger = U \xi R^\dagger,
\label{xiT}
\end{equation}
where $U$ is a complicated function of $L,R$, and the pion fields $\Pi$.   In the special case where $L = R = V$ in the unbroken $SU(2)_V$ vector subgroup, 
$U = V$.

We present an effective Lagrangian for the strong interactions of low momentum pions (in case of $SU(3)_L \otimes SU(3)_R$, these results will include kaons 
and the eta, as well) with the ground state heavy mesons with $Q \bar q^a$  flavor quantum numbers, where $a = 1, 2$, and $q^1 = u$, $q^2 = d$.  The light degrees of freedom have $s_l^{\pi_l} = {1\over2} ^-$ spin-parity quantum numbers, in these heavy mesons.  In the limit where the mass of the heavy quark $m_Q \to \infty$, the spin of the light degrees of freedom combines with the spin of the heavy quark to yield two degenerate doublets, consisting of an $SU(2)_V$ antidoublet of pseudo-scalar mesons, denoted by $P_a$, and an $SU(2)_V$ antidoublet of vector mesons, denoted by $P_a^*$.  We are interested in the case $Q = c$, for which the pseudo-scalar mesons are $D^0$ and $D^+$, 
and the vector mesons are $D^{*0}$ and $D^{*+}$.  The above mentioned strong 
interaction Lagrangian, in addition to the usual symmetries, such as parity 
and Lorentz invariance, must have heavy quark symmetry, at the leading order.  To proceed, it is convenient to incorporate the $P_a$ and $P_{a\mu}^*$ meson fields into a $4 \times 4$ matrix $H_a$ \cite{Rapid, Rapid 14}
\begin{equation}
H_a = {{1 + \not\! v}\over2}(P_{a\mu}^* \gamma^\mu - P_a\gamma_5).
\label{H}
\end{equation}
Note that the heavy fields $P_a$ and $P_{a\mu}^*$ only destroy their respective mesons of four-velocity $v$ and do not create the corresponding antiparticles.  We have $v^\mu P_{a\mu}^* = 0$.  Under $SU(2)_L \otimes SU(2)_R$
\begin{equation}
H_a \to H_b U_{ba}^\dagger,
\label{HT}
\end{equation}
where the repeated index $b$ is summed over 1 and 2, and $U$ was introduced in Eq. (\ref{xiT}).
Under the heavy quark spin symmetry group $SU(2)_v$, we have
\begin{equation}
H_a \to S H_a,
\label{HS}
\end{equation}
where $S\in SU(2)_v$.
Lorentz transformations act on $H_a$ according to
\begin{equation}
H_a \to D(\Lambda) H_a D(\Lambda)^{-1},
\label{HL}
\end{equation}
where $D(\Lambda)$ is an element of the $4\times4$ matrix representation of the Lorentz group.

We introduce
\begin{equation}
\bar H_a = \gamma^0 H_a^\dagger \gamma^0.
\label{Hbar}
\end{equation}
Thus, we get
\begin{equation}
\bar H_a = (P_{a\mu}^{*\dagger} \gamma^\mu + P_a^\dagger \gamma_5)
{{1 + \not\! v}\over2}.
\label{HbarExp}
\end{equation}
The transformation laws for $\bar H_a$ corresponding to those in Eqs. 
(\ref{HT}), (\ref{HS}), and (\ref{HL}), are
\begin{equation}
\bar H_a\to U_{ab}\bar H_b, \bar H_a \to \bar H_a S^{-1}, ~{\rm and}~ \bar H_a \to  D(\Lambda) \bar H_a D(\Lambda)^{-1}.
\label{HbarT}
\end{equation}

The effective Lagrangian that describes the strong interactions of pions and heavy mesons, at leading order (one derivative), is then given by
\begin{equation}
{\cal L}^{(H)} = -i Tr \bar H_a v \cdot \partial H_a + i Tr \bar H_a H_b{(v\cdot V)}_{ba} - g_D Tr \bar H_a H_b \gamma_\lambda \gamma_5 {(A^\lambda)}_{ba} ,
\label{LH}
\end{equation}
where 
\begin{equation}
V_\nu = {1\over 2} (\xi \partial_\nu \xi^\dagger + \xi^\dagger \partial_\nu
\xi)
\label{V}
\end{equation}
and
\begin{equation}
A_\lambda = {i\over 2} (\xi\partial_\lambda \xi^\dagger - \xi^\dagger
\partial_\lambda \xi).
\label{A}
\end{equation}

At the present time, there is only a bound on the coupling constant $g_D$ of Eq. (\ref{LH}), $g_D^2 < 0.45$, coming from the experimental bound $\Gamma (D^{*+} \to D^0 \pi^+) < 0.09$ MeV \cite{PDG}.
The explicit chiral symmetry breaking terms contain light quark masses
\begin{equation}
\delta {\cal L}^{(1)} = \lambda_\xi Tr \bar H_b H_a {(M_\xi)}_{ab} + \lambda_\Sigma Tr \bar H_a H_a {(M \Sigma + \Sigma^\dagger M)}_{bb}.
\label{L1}
\end{equation}
In Eq. (\ref{L1}),
\begin{equation}
M_\xi = {1\over 2} (\xi M\xi + \xi^\dagger M \xi^\dagger),
\label{Mxi}
\end{equation}
where $M = diag (m_u, m_d)$ is the $2 \times 2$ light quark mass matrix.

Eq. (\ref{LH}) yields the propagators $i\delta_{ab}/2v \cdot k$ and $-i\delta_{ab}(g^{\mu \nu} - v^\mu v^\nu)/2v \cdot k$ for $P_a$ and $P_a^*$, respectively, where $k$ is a small residual momentum.  The leading heavy quark spin symmetry breaking effects at order 
$\Lambda_{QCD}/m_Q$, induced by the color-magnetic operator \cite{Rapid 6}, 
are given by
\begin{equation}
\delta {\cal L}^{(2)} = {\lambda_Q \over m_Q} Tr \bar H_a \sigma^{\mu \nu} H_a \sigma_{\mu \nu}.
\label{L2}
\end{equation}
In Eq. (\ref{L1}), $\lambda_\xi$ and $\lambda_\Sigma$ are dimensionless constants independent of the heavy quark mass, whereas $\lambda_Q$ of Eq. (\ref{L2}), with dimension 2, has a logarithmic dependence on the heavy quark mass \cite{Rapid 6}, calculable in perturbative QCD.  Under $SU(3)_L \otimes SU(3)_R$, the correspondence with the notation of Ref. \cite{Rapid} is 
given by 
\begin{equation}
g_D = g, \lambda_\xi = 2\lambda_1, \lambda_\Sigma = \lambda_1^\prime, 
~{\rm and}~ \lambda_Q = \lambda_2.
\label{gD}
\end{equation}

The only effect of the term in Eq. (\ref{L2}) is to change the $P_a$ and 
$P_a^*$ propagators, and an appropriate field redefinition \cite{Rapid}
will yield $i\delta_{ab}/2v \cdot k$ and $-i\delta_{ab}(g^{\mu \nu} - v^\mu v^\nu)/2(v \cdot k - \Delta)$  for the aforementioned propagators, respectively,
where 
\begin{equation}
\Delta = m_{P^*} - m_P = {-2\lambda_Q \over m_Q}.
\label{Delta}
\end{equation}
In this paper $Q=c$, and the mass difference $\Delta = m_{D^*} - m_D = 145$ MeV
$\sim m_\pi$.  Thus, $\Delta$ is considered as of order one derivative and has 
a leading order contribution in our subsequent calculations.

The $K^*$ and $\bar K^*$ fields are introduced as doublets \cite{D.W.}
\begin{equation}
K^*_\mu = \left[ \begin{array}{c}
K^{*+}_\mu\\
K^{*0}_\mu
\end{array} \right], \qquad \bar K_\mu^* = \left[ \begin{array}{c}
K^{*-}_\mu\\
\bar K^{*0}_\mu
\end{array} \right] .
\label{K*}
\end{equation}
Under chiral $SU(2)_L \otimes SU(2)_R$,
\begin{equation}
K_\mu^* \rightarrow U K_\mu^*,~~
\bar K_\mu^* \rightarrow U^* \bar K^*_\mu .
\label{K*T}
\end{equation}
The doublets $K^*_\mu$ and $\bar K_\mu^*$ are related by charge conjugation $C$
which acts on the fields as follows:
\begin{equation}
C K_\mu^* C^{-1} = - \bar K_\mu^*,\quad C\xi
C^{-1} = \xi^T.
\label{K*C}
\end{equation}

The vector meson fields are treated as heavy with fixed four velocity 
$v^{\prime \mu}$, $v^{\prime 2} = 1$, satisfying the constraint $v^\prime \cdot K^* = v^\prime \cdot \bar K^* = 0$.  Interactions of the form $V \rightarrow V' X$, where $V$ and $V'$ are $K^*$'s and $X$ is either the vacuum or one or more soft pions, are given by a Lagrangian of the form 
\begin{equation}
{\cal L} = {\cal L}_{kin} + {\cal L}_{int} + {\cal L}_{mass} - {i\over 2} {\cal
L}_{width} .
\label{L}
\end{equation}

The interaction Lagrangian is given by
\begin{equation}
{\cal L}_{int} = ig_2^{(K^*)} \bar K_\mu^{*\dagger} A_\lambda^T \bar K_\nu^* 
v_ \sigma^\prime \epsilon^{\mu\nu\lambda\sigma}
+ig_2^{(K^*)} K_\mu^{*\dagger} A_\lambda K_\nu^* v_\sigma^\prime
\epsilon^{\mu\nu\lambda\sigma}.
\label{Lint}
\end{equation}
Here, $g_2^{(K^{*})}$ in Eq. (\ref{Lint}) and $g_2$ of the Lagrangian in Eq. (11) of Ref. \cite{Jenkins} are equal, at the leading order in $SU(3)_L \otimes SU(3)_R$ chiral perturbation theory.  Note that for the vector mesons $K_\mu^{-*\dagger} \not= K_\mu^{*+}$, and so on.  In the heavy vector meson chiral perturbation theory, $K_\mu^{*+}$ destroys a $K^{*+}$, but it does not create its antiparticle $K^{-*}$.  The field $K_\mu^{-*\dagger}$ creates a $K^{*-}$.

The kinetic terms in Lagrangian (\ref{L}) are 
\begin{equation}
{\cal L}_{kin}  = 
- i K_\mu^{*\dagger} v^\prime \cdot \partial K^{*\mu} - i K_\mu^{*\dagger} v^\prime \cdot V K^{*\mu} - i \bar K_\mu^{*\dagger} v^\prime \cdot \partial 
\bar K^{*\mu} + i \bar K_\mu^{*\dagger} v^\prime \cdot V^T \bar K^{*\mu}.
\label{Lkin}
\end{equation}

The mass terms which explicitly break chiral symmetry, are given by 
\[
{\cal L}_{mass} = \lambda_2^{(K^{*})} K_\mu^{*\dagger} M_\xi K^{*\mu}
+\lambda_2^{(K^{*})} \bar K_\mu^{*\dagger} M_\xi^T \bar
K^{*\mu}
+\sigma_8^{(K^{*})} Tr (M_\xi) K_\mu^{*\dagger} K^{*\mu} \]
\begin{equation}
+\sigma_8^{(K^{*})} Tr (M_\xi) \bar K_\mu^{*\dagger} \bar K^{*\mu}.
\label{Lmass}
\end{equation}
At leading order in $SU(3)_L \otimes SU(3)_R$ chiral perturbation theory, the couplings in Eq. (\ref{Lmass}) are related to those of Ref. \cite{Jenkins} by
\begin{equation}
\lambda_2^{(K^{*})} = \lambda_2 ~{\rm and}~
\sigma_8^{(K^{*})} = \sigma_8.
\label{ls}
\end{equation}

The $K^*$'s have a width  $\Gamma^{(K^{*})}
= 50$ MeV.  Since this width is compararble to the pion mass, we treat it as
of order one derivative, and introduce it in our Lagrangian via the following terms
\begin{equation}
{\cal L}_{width} =\Gamma^{(K^{*})} K_\mu^{*\dagger} K^{*\mu}
+\Gamma^{(K^{*})} \bar K_\mu^{*\dagger} \bar K^{*\mu}.
\label{Lwidth}
\end{equation}
Note that the terms in Eqs. (\ref{Lint}), (\ref{Lkin}), and (\ref{Lwidth}) are of order one derivative and are considered of leading order, whereas the terms in Eq. (\ref{Lmass}) are proportinal to light quark masses, which means they 
are of order two derivatives, and thus, considered non-leading in our calculations.  In this formalism, the $K^*$ propagator is given by  
\begin{equation}
{-i (g^{\mu\nu} - v^{\prime \mu} v^{\prime \nu}) \over v^\prime \cdot k^\prime 
+ i\Gamma^{(K^{*})}/2},
\label{PropK*}
\end{equation}
where $k^\prime$ is the small residual momentum of the $K^*$.
\section{Currents and the Amplitude}

The part of the effective Hamiltonian ${\cal H}_W$ for weak semileptonic decay of $D^0$ that contributes to $D^0\rightarrow \bar K^{*0} \pi^- e^+ \nu_e$, at the quark 
level, is given by
\begin{equation}
{\cal H}_W = {G_F\over \sqrt 2} V_{cs} \bar \nu_e \gamma_\mu (1-\gamma_5)e^+
\bar s\gamma^\mu (1-\gamma_5)c,
\label{Hw}
\end{equation}
where $G_F$ is the Fermi constant, $V_{cs}$ is an element of the Cabibbo-Kobayashi-Maskawa matrix, and the spinors reprsent the corresponding 
fermions.  Experimentally, we have $|V_{cs}| \approx 1$.  The only Feynman diagrams that contribute at the leading order in $SU(2)_L \otimes SU(2)_R$ chiral perturbation theory are presented in Figs. (1) and (2).
\begin{figure}[htbp]
\centerline{\epsfxsize=8truecm \epsfbox{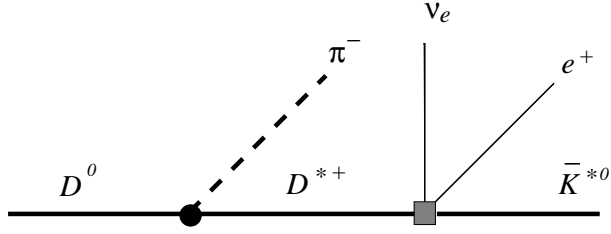}}
\caption[1]{The $D^*$-pole diagram contribution to the amplitude in Eq. (\ref{AMP}).  The solid square represents the hadronic left-handed current and the solid circle represents the $DD^*\pi$ coupling proportional to $g_D$. }
\end{figure}

\newpage
We need expressions for $\langle K^*|\bar s\gamma^\mu (1-\gamma_5)c|D^*\rangle$ and
$\langle K^*|\bar s\gamma^\mu (1-\gamma_5)c|D\rangle$, in order to write down the amplitude for the decay.  We write \cite{Isgur} 
\begin{equation}
\langle K^*| (p^\prime, \varepsilon^\prime)|V_\mu|D(p)\rangle 
\equiv ig\epsilon_{\mu\nu\lambda\sigma} \varepsilon^{*\prime\nu}
(p + p^\prime)^\lambda (p - p^\prime)^\sigma
\label{Vmu}
\end{equation}
and
\begin{equation}
\langle K^*| (p^\prime, \varepsilon^\prime)|A_\mu|D(p)\rangle \equiv 
f\varepsilon^{*\prime}_\mu + a^{(+)}(\varepsilon^{*\prime}\cdot p)(p + p^\prime)_\mu + a^{(-)}(\varepsilon^{*\prime}\cdot p)(p - p^\prime)_\mu,
\label{Amu}
\end{equation}
where $p^{\prime}$ and $\varepsilon^{\prime}$ are the four-momentum and the polarization of the $K^*$, $p$ is the four-momentum of the $D$, $V_\mu\equiv
\bar s \gamma_\mu c$, $A_\mu\equiv \bar s\gamma_\mu\gamma_5 c$, and $g, f, a^{(+)}$, and $a^{(-)}$ are experimentally measurable form factors which are functions of $p \cdot p^{\prime}$.  At the present time, $a^{(-)}$, having a 
contribution to $D\to K^* \bar \ell \nu_\ell$ that is proportional to the 
lepton mass, remains unmeasured, but $g, f$, and $a^{(+)}$ have been measured
\cite{PDG}.  Let $p=m_Dv$ and $p^{\prime}=m_{K^*}v^{\prime}$, where $v$ and $v^{\prime}$ are the four-velocities of the $D$ and the $K^*$, respectively.  Then, as a function of $y\equiv v\cdot v^{\prime}$, we have 
\cite {Ligeti}
\begin{equation}
f(y)={1.8~{\rm GeV}~\over 1+0.63(y-1)},
\label{f}
\end{equation}
\begin{equation}
a^{(+)}(y)=-{0.17~{\rm GeV}~^{-1}\over 1+0.63(y-1)},
\label{a+}
\end{equation}
and
\begin{equation}
g(y)={0.51~{\rm GeV}~^{-1}\over 1+0.96(y-1)}.
\label{g}
\end{equation}
Note that the sign of $g$ depends on the convention for the sign of Levi-Civita tensor, which we take to be $\epsilon_{\mu\nu\lambda\sigma}=-\epsilon^{\mu\nu\lambda\sigma}=1$.
\begin{figure}[tbp]
\centerline{\epsfxsize=8truecm \epsfbox{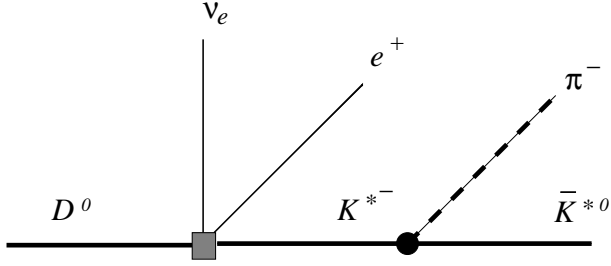}}
\caption[1]{The $K^*$-pole diagram contribution to the 
amplitude in Eq. (\ref{AMP}).  The solid square represents the hadronic left-handed current and the solid circle represents the $K^*K^*\pi$ coupling proportional to $g_2^{(K^*)}$.}
\end{figure}

Next, we will write an expression for $\langle K^*|\bar s \gamma_\mu(1-\gamma_5)c|D^*\rangle$, in terms of the form factors $g, f, a^{(+)}$, and $a^{(-)}$.  Later on, we will only consider the region of phase space where $\bar K^{*0}$ is nearly at rest in the decaying $D^0$ rest frame 
and the dependence of the amplitude on the unknown form factor $a^{(-)}$ is negligible.  We treat the charm quark as heavy and use heavy quark symmetry to write down 
the following expression \cite {Zoltan} for the current $\langle K^*|V_\mu-A_\mu|M^{(H)}\rangle$, where $M^{(H)}$ is either a $D$ or a $D^*$ with polarization $\varepsilon$
\begin{equation}
\langle K^*|V_\mu-A_\mu|M^{(H)}\rangle=Tr\left[K^*(v, v^{\prime}, \varepsilon^{\prime})\gamma_\mu
(1-\gamma_5)M^{(H)}(v)\right].
\label{Jmu}
\end{equation}
In Eq. (\ref{Jmu}),  
\begin{equation}
M^{(H)}(v) = {1+\not\! v \over 2}\left\{ \begin{array}{ll}
-\gamma_5  & \mbox{; \, for $D$} \\
\not\!\varepsilon  & \mbox{; \, for $D^*$},
\end{array}
\right.
\label{M}
\end{equation}
and 
\begin{equation}
K^*(v,v^{\prime},\varepsilon^{\prime}) = (C_1+C_2\not\! v^{\prime})\not\!\varepsilon^{*\prime}+(C_3+C_4\not\! v^{\prime})(\varepsilon^{*\prime}\cdot v)
\label{K*vv'}
\end{equation}
is the most general form that can be written down for the wavefunction of 
$K^*$, consistent with heavy quark symmetry for $M^{(H)}(v)$, at the leading order.
$C_i, i=1, 2, 3, 4$, are form factors which can be expressed in terms of $f, 
a^{(+)}, a^{(-)}$, and $g$, by comparing the expression for $\langle K^*|V_\mu-A_\mu|D\rangle$ from Eqs. (\ref{Vmu}) and (\ref{Amu}) with the one given 
by Eq. (\ref{Jmu}) above.  With the $C_i$ determined in this way, Eq. (\ref{Jmu}) yields the following expression for $\langle K^*|V_\mu-A_\mu|D^*\rangle$, in terms of 
the form factors of $\langle K^*|V_\mu-A_\mu|D\rangle$
\begin{eqnarray}
&&\langle K^*(v^\prime,\varepsilon^\prime)|V_\mu - 
A_\mu|D^*(v,\varepsilon)\rangle = 
m_Dm_{K^*}\Bigg(
i\epsilon_{\mu\nu\lambda\sigma}\varepsilon^\nu \bigg[(a^{(+)} -a^{(-)})
v^\lambda v^{\prime\sigma}(\varepsilon^{*\prime} \cdot v)\nonumber\\
&&- {f \over m_Dm_{K^*}}\varepsilon^{*\prime\lambda}v^\sigma \bigg]+\bigg[{f \over m_Dm_{K^*}}
+ {m_D \over m_{K^*}}(a^{(+)} + a^{(-)}) 
+(a^{(+)} - a^{(-)})(v \cdot v^{\prime})\bigg](\varepsilon^{*\prime} \cdot v)\varepsilon_\mu \nonumber\\
&&+ 2g(\varepsilon \cdot v^{\prime})\varepsilon^{*\prime}_\mu 
+\bigg\{2g\big[i\epsilon^{\alpha\beta\gamma\delta}\varepsilon_\alpha
v_\beta\varepsilon^{*\prime}_\gamma v^\prime_\delta + (\varepsilon
\cdot \varepsilon^{*\prime})(v \cdot v^{\prime}) -(\varepsilon \cdot v^{\prime})(\varepsilon^{*\prime} \cdot v)\big] \nonumber\\
&&-(a^{(+)} - a^{(-)})(\varepsilon \cdot v^{\prime})(\varepsilon^{*\prime} 
\cdot v)
-{f \over m_Dm_{K^*}}(\varepsilon \cdot \varepsilon^{*\prime})\bigg\}v_\mu 
-2g(\varepsilon \cdot \varepsilon^{*\prime})v^{\prime}_\mu\Bigg),
\label{JD*K*}
\end{eqnarray}
where we have used the identity $\varepsilon_\alpha \varepsilon^{*\prime}_\beta
v^\prime_\gamma v_\lambda v_\sigma g^{\lambda [ \sigma }
\epsilon^{\alpha \beta \gamma \mu ]}=0$, to write down the first term 
proportional to $v_\mu$.

We see that the expressions for $\langle K^*|V_\mu-A_\mu|D\rangle$ and $\langle K^*|V_\mu-A_\mu|D^*\rangle$, given by Eqs. (\ref{Vmu}), (\ref{Amu}), and (\ref{JD*K*}) depend on the value of $a^{(-)}$, which, as mentioned before, is not measured.  The value of this form factor could, in principle, be measured, given enough data on $D \to K^* \mu^+ \nu_\mu$, where the anti-muon is massive enough to make the measurement possible.  However, for the rest of this paper, we work in the kinematic regime where $\bar K^{*0}$ is at rest, to leading order in our calculations, in the rest frame of the decaying $D^0$.  Thus, the pion can be soft in the rest frames of both the $D^0$ and the $\bar K^{*0}$, as required by chiral perturbation theory.  In this region, we have $v=v^\prime$, $v \cdot v^\prime=1$, and $\varepsilon \cdot v^{\prime}=\varepsilon^{*\prime} \cdot v=0$.  The amplitude for the decay $D^0\rightarrow \bar K^{*0} \pi^- e^+ \nu_e$, at leading order in $SU(2)_L \otimes SU(2)_R$ chiral perturbation theory, represented by Feynman diagrams of Figs. (1) and (2), is given by
\[
{\cal M}(D^0\to \bar K^{*0} \pi^- e^+ \nu_e) = {G_F V_{cs} f \over 
\sqrt 2 f_\pi}\bigg\{ {g_D \over v\cdot p_\pi+
\Delta}\big[\epsilon^{\mu\nu\lambda\sigma}v_\nu p_{\pi\lambda}
\varepsilon^{*\prime}_\sigma -i(p_\pi \cdot \varepsilon^{*\prime})v^\mu \big] 
\]
\begin{equation}
+{g_2^{(K^*)} \over v\cdot p_\pi + i(\Gamma^{(K^*)} / 2)}  \epsilon^{\mu\nu\lambda\sigma}p_{\pi\nu}\varepsilon^{*\prime}_\lambda v_\sigma
\bigg\} \bar u_{(\nu)} \gamma_\mu(1-\gamma_5)v_{(e)},
\label{AMP}
\end{equation}
where $u_{(\nu)}$ and $v_{(e)}$ are the spinors for the electron neutrino
and the positron, respectively.  Note that in the recoilless $\bar K^{*0}$
limit used here, the dependence of the amplitude on $a^{(-)}$ is eliminated.  
We have $f(1)=1.8$ GeV, from Eq. (\ref{f}) above.

\section{Differential Decay Rate}

Before presenting the differential decay rate, let us introduce the following kinematic variables \cite {Lee,Lee8}.  The invariant mass of the $K^*\pi$ system $m_{K^*\pi}$, where $m_{K^*\pi}\equiv \sqrt{(p^\prime + p_\pi)^2}$, is given by
\begin{equation} 
m_{K^* \pi}^2=m_{K^*}^2 + m_\pi^2 + 2m_{K^*}m_\pi x,
\label{mkp}
\end{equation}
where 
\begin{equation}
x \equiv {v \cdot p_\pi \over m_\pi}.
\label{x}
\end{equation}

The invariant mass of the lepton pair is denoted by $m_{e\nu}$, where 
$m_{e\nu} \equiv \sqrt{(p_e + p_\nu)^2}$, $p_e$ is the four-momentum of the 
positron, and $p_\nu$ is the four-momentum of the electron neutrino.  The angle formed by the three momentum of the $\bar K^{*0}$ in the $K^* \pi$ center of mass frame and the line of flight of the $K^* \pi$ in the $D^0$ rest frame is denoted by $\theta_K$, and the angle between the positron three momentum in the 
$e^+ \nu$ center of mass frame and the line of flight of the $e^+ \nu$ in the $D^0$ rest frame is denoted by $\theta_e$.  The angle $\phi$ is formed by the normals to the $K^* \pi$ and $e^+ \nu$ planes in the $D^0$ rest frame.  The sense of $\phi$ is from the normal to the $K^* \pi$ plane to that of the  $e^+ \nu$ plane.

The amplitude given by Eq. (\ref{AMP}) is obtained in a region where the $\bar K^{*0}$ is ``near zero recoil'', and later we specify a region of phase space that corresponds to this approximation.  Using the above kinematic variables, 
in the limit $m_\pi/m_{K^*} \to 0$, one obtains the following differential 
decay rate for $D^0 \to \bar K^{*0} \pi^- e^+ \nu_e$, in a region where $\bar K^{*0}$ is near zero recoil
\begin{eqnarray}
&&{d^{(5)}\Gamma(D^0 \to \bar K^{*0} \pi^- e^+ \nu_e) \over dxdm_{e\nu}^2 d(\cos\theta_K)d(\cos\theta_e)d\phi} =
{G_F^2|V_{cs}|^2 f^2(1) \over (4\pi)^6 f_\pi^2 m_D^3}\left({m_\pi \over
m_{K^*}}\right)^2 (x^2 - 1)^{3/2} \sqrt{T_0} \nonumber\\
&&\times\Bigg\{{g_D^2 \over (x+\delta)^2 } (T_1 + T_2) 
+\bigg[{{g_2^{(K^*)}}^2 \over x^2 + \gamma^2}  
+ {2 x \, g_D\, g_2^{(K^*)}  \over
(x+\delta)(x^2+\gamma^2)}\bigg] T_2 \Bigg\} ,
\label{d5G}
\end{eqnarray}
where
\begin{equation}
\gamma \equiv {\Gamma^{(K^*)} \over 2 m_\pi} \,\,\,\, ; \,\,\,\, \delta \equiv {\Delta \over
m_\pi},
\label{gd}
\end{equation}
\begin{equation}
T_0 \equiv (m_D^2 + m_{K^*}^2 -m_{e\nu}^2)^2  - 4m_D^2 m_{K^*}^2,
\label{T0}
\end{equation}
\begin{equation}
T_1 \equiv X^2 \sin^2 \theta_e,
\label{T1}
\end{equation}

\begin{eqnarray}
T_2 &\equiv& \sin^2\theta_K \sin^2\theta_e (X^2 + m_{K^*}^2m_{e\nu}^2 \cos^2\phi
)+m_{K^*}^2m_{e\nu}^2(1+\cos^2\theta_K \cos^2\theta_e) \nonumber\\
&-&\left({m_{K^*}m_{e\nu} \over 2}\right) \sin(2\theta_K) \sin(2\theta_e) \cos\phi
\sqrt{X^2 + m_{K^*}^2m_{e\nu}^2}\, ,
\label{T2}
\end{eqnarray}
and
\begin{equation}
X \equiv \sqrt {{(m_D^2 - m_{K^*}^2 -m_{e\nu}^2)^2 \over 4} - m_{K^*}^2 m_{e\nu}^2}\, .
\label{X}
\end{equation}

We can integrate the expression in Eq. (\ref{d5G}) for the differential decay rate,  in order to obtain a total decay rate $\Gamma(D^0 \to \bar K^{*0} 
\pi^- e^+ \nu_e)$, for a limited volume of phase space consistent with the regime of validity of chiral perturbation theory and the recoilless $\bar K^{*0}$ approxiamtion.  In this region of phase space, we demand that

(a)  $p \cdot p_\pi / m_D \ll \Lambda_{\chi SB}$ and  $p^\prime \cdot p_\pi / m_{K^*} \ll \Lambda_{\chi SB}$, where $\Lambda_{\chi SB} \sim 1$ GeV, in oredr to have a valid perturbative expansion, and

(b)  $\left|\vec {p^\prime}\right| / m_{K^*} \ll 1$, where $\vec {p^\prime}$ 
is the three momentum of the $\bar K^{*0}$, in the rest frame of the $D^0$, 
to ensure that we stay in a region of phase space where $v = v^\prime$,  corresponding to a recoilless $\bar K^{*0}$, at leading order.

The constraints in (a) and (b) above are not very exact and an appropriate 
choice for the region of phase space that satisfies our requirements may, in principle, be made only after comparing the shapes of different distributions 
from experimental data with the predictions from the theoretical differential 
decay rate presented in Eq. (\ref{d5G}).  The possibility of making such comparisons between experimental data and our theory depends on the 
availability and resolution of the data in the restricted phase space region 
of our calculations. 

However, we proceed to make a reasonable choice for the region of integration,
given the constraints mentioned in (a) and (b) above.  To satisfy  $p^\prime \cdot p_\pi / m_{K^*} \ll \Lambda_{\chi SB}$, we require $x\in [1, 2]$.  The quantities $p \cdot p_\pi / m_D$ and $\left|\vec {p^\prime}\right|$ depend
on the values of the variables $x$, $m_{e\nu}$, and $\cos \theta_{K}$, as 
shown in the appendix.  By inspection, for $x\in [1, 1.5]$, 
$\cos \theta_K \in [-1, 1]$, and 
$m_{e\nu}^2 \in [0.593, (m_D - m_{K^* \pi})^2]$, we have
$p \cdot p_\pi / m_D < 200$ MeV and $\left|\vec {p^\prime}\right| < 240$ MeV.  Integrating the expression in Eq. (\ref{d5G}) over $m_{e\nu}^2 \in [0.593, 
(m_D - m_{K^* \pi})^2]$, $x\in [1, 1.5]$, $\cos \theta_K \in [-1, 1]$, $\cos \theta_e \in [-1, 1]$, and $\phi \in [0, 2\pi]$, yields the partial decay 
width  $\Gamma_1(D^0 \to \bar K^{*0} \pi^- e^+ \nu_e)$, and
\begin{equation}
\Gamma_1(D^0 \to \bar K^{*0} \pi^- e^+ \nu_e) = 9.01\times10^{-17}\big[(0.016)g_D^2 \pm (0.024)g_D + 0.013 \big] ~{\rm GeV}~,
\label{G1}
\end{equation}
where $\pm$ corresponds to $g_2^{(K^*)} = \pm 0.6$.
With the angular limits of integration the same, if we assume that, without exceeding the realm of validity of our theory considerably, the 
region of integration may be expanded to $x\in [1, 2]$ 
and $m_{e\nu}^2 \in [0.510, (m_D - m_{K^* \pi})^2]$, where 
$p \cdot p_\pi / m_D < 245$ MeV and $\left|\vec {p^\prime}\right| < 330$ MeV, 
we obtain a partial decay width $\Gamma_2(D^0 \to \bar K^{*0} \pi^- e^+ \nu_e)$, which is nearly an order of 
magnitude larger
\begin{equation}
\Gamma_2(D^0 \to \bar K^{*0} \pi^- e^+ \nu_e) = 9.01\times10^{-17}\big[(0.176)g_D^2 \pm (0.231)g_D + 0.117 \big] ~{\rm GeV}~,
\label{G2}
\end{equation}
where, again, $\pm$ corresponds to $g_2^{(K^*)} = \pm 0.6$.
We mentioned before that present data suggests $g_D^2<0.45$.  Let us take 
$g_D^2 = 0.3$ in order to get some estimates on the size of the branching 
ratio in our region of phase space.  Since the total width $\Gamma_{D^0} = 1.586\times10^{-12}$GeV for the $D^0$, $\Gamma_1$ gives a branching 
ratio $B_1$ 
\begin{equation}
B_1(D^0 \to \bar K^{*0} \pi^- e^+ \nu_e)=\left\{\begin{array}{ll}
2\times10^{-6} & \mbox{;\, $g_Dg_2^{(K^*)}>0$}\\
3\times10^{-7} & \mbox{;\, $g_Dg_2^{(K^*)}<0,$}
\end{array}
\right.
\label{B1}
\end{equation}
in a region where $m_{e\nu}^2 \in [0.593, (m_D - m_{K^* \pi})^2]$ and 
$x\in [1, 1.5]$.  For the branching ratio $B_2$, corresponding to $\Gamma_2$, 
we get 
\begin{equation}
B_2(D^0 \to \bar K^{*0} \pi^- e^+ \nu_e)=\left\{\begin{array}{ll}
2\times10^{-5} & \mbox{;\, $g_Dg_2^{(K^*)}>0$}\\
3\times10^{-6} & \mbox{;\, $g_Dg_2^{(K^*)}<0$}
\end{array}
\right.
\label{B2}
\end{equation}
in our restricted region of phase space, where $m_{e\nu}^2 \in [0.510, 
(m_D - m_{K^* \pi})^2]$ and $x\in [1, 2]$.  For the smaller value of $B_1$, the corresponding region can most likely be explored at a fixed target experiment, or a tau-charm or $B$ factory only, since the present experiments are not able to measure such 
small branching ratios for the decay of the $D^0$.  However, the values of 
$B_2$ lie reasonably close to the present measureable range, and $g_D$ can be extracted from the data if its absolute value is not too much smaller than $\sqrt {0.3}$.

\section{Conclusions}

In this work, we have presented a systematic calculation of the differential decay rate for $D^0 \to \bar K^{*0} \pi^- e^+ \nu_e$, in a restricted kinematic region, based on the $SU(2)_L \otimes SU(2)_R$ chiral perturbation theory formalism.  Since this decay rate depends on the unmeasured $DD^*\pi$ coupling constant $g_D$ of Eq. (\ref{LH}), one can use our results to extract $g_D$ from experimental data.  Currently, there is an experimental bound, $g_D^2<0.45$, 
on this coupling constant.

We have treated $D$, $D^*$, and $K^*$ as heavy matter fields and applied the chiral perturbation theory formalism to describe the strong couplings of the $\pi^-$ to $D$, $D^*$, and $K^*$, assuming that the $\pi^-$ is ``soft'' in the rest frames of the heavy matter fields.  The leading order (one derivative, here) amplitude involves $\langle K^* |\bar s \gamma_\mu
(1-\gamma_5)|D\rangle$ and $\langle K^* |\bar s \gamma_\mu(1-\gamma_5)|D^*\rangle$ left-handed hadronic currents, and we have presented an expression, in Eq. (\ref{JD*K*}), for the $D^* \to K^*$ current, 
in terms of the form factors of the $D \to K^*$ current.  One of the $D \to K^*$  form factors, denoted by $a^{(-)}$, remains unmeasured. However, we have restricted our phase space to a region where the $\bar K^{*0}$ is near zero recoil in the $D^0$ rest frame, making the dependence of the amplitude on $a^{(-)}$ negligible, at the leading order.  Note that this restriction on the phase space is made in accordance with the requirement that the pion be soft in the rest frames of both the $D^0$ and the $\bar K^{*0}$, in order to have a valid chiral perturbation theory expansion.  

We have presented an expression for the differential decay rate,
in the aforementioned region of phase space, in Eq. (\ref{d5G}).  To get the branching ratio in this region, one has to make a reasonable choice for the limits of the integration that yields the partial width in the restricted 
volume of phase space.  For $|g_D| = \sqrt {0.3}$ as an allowed value for the magnitude of the $DD^*\pi$ coupling, we get the branching ratios, $B_1$ and $B_2$, corresponding to two reasonable choices for the integration region.  The first choice is expected to be strict enough for a good leading order approximation, and it can probably be explored at a fixed target experiment, or a tau-charm or $B$ factory only.  We believe that expanding the region of integration beyond that which corresponds to the second choice can result in a considerable departure from the proper regime for our approximations.  The branching ratios in this region are close to the present measureable range.  For a rough experimental value of $g_D$ we can expect that
the branching ratio for a typical region of integration be of order $10^{-5}$.

The volume of phase space in which our theory is valid may be best selected after consulting the data.  However, the values of $B_2$ suggest that for a volume of phase space that is close the our second choice presented here, and provided that $|g_D|$ is not too much smaller than the value used here, $\sqrt {0.3}$, an experimental value for $g_D$ can be extracted, using the data on  $D^0 \to \bar K^{*0} \pi^- e^+ \nu_e$, and given the sign 
of $g_2^{(K^*)}$.  In the heavy quark limit ($m_Q \to \infty$), $g_D$ is equal to $g$ in Eq. (12) of Ref. \cite{Rapid}.  There are other processes, with 
fewer final state particles, that involve the constant $g_D$.  $D^* \to D \pi$ is an example of such a process, where one, in principle, can measure the 
value of $g_D$, from the knowledge of the partial width for this decay 
channel.  This assumes knowing the full width of $D^*$, which is difficult to measure, since the $D^*$ is too short lived for time of flight measurements, 
and too narrow for a reliable measurement of its full width.

Our results represent the $SU(2)_L \otimes SU(2)_R$ chiral perturbation thoery predictions at the level of one derivative, corresponding to ${\cal O}
(v_H \cdot p_\pi/1{\rm GeV})$, where $v_H$ is the four-velocity of the $D^0$ 
or $\bar K^{0*}$.  Thus, we expect to have significant corrections to our 
results at the level of two derivatives, supressed by one power of $v_H \cdot p_\pi/1{\rm GeV}$, only.  Hence, at the next to leading order, operators containing two derivatives of the pion field, insertions of the light quark 
mass terms, and additional pole graphs, such as the one in which the $K^{-*}$
is replaced by a $K^-$ in Fig. 2, contribute.

\section*{Acknowledgements}
The author wishes to extend his sincere gratitude to Mark Wise for providing invaluable guidance throughout this work.  It is a pleasure to thank Anton Kapustin and Zoltan Ligeti for bringing some of the formalism used in this work to my attention.  Peter Cho, John Elwood, Martin Gremm, Adam Leibovich, Jon Urheim, and Eric Westphal are also thanked for helpful discussions.
\newpage

\section*{Appendix}

The following relations \cite{Lee} are useful in calculating the scalar products of the varoius four-vectors introduced in this paper.  Let
\[
R\equiv p^{\prime} + p_\pi \,\,\,\, ; \,\,\,\, Q\equiv p^{\prime} - p_\pi 
\]
\begin{equation}
K\equiv p_e + p_\nu \,\,\,\, ; \,\,\,\, L\equiv p_e - p_\nu.
\label{dot1}
\end{equation}
We then get
\[
Q \cdot R = m_{K^*}^2 - m_\pi^2
\]
\[
K \cdot L=0
\]
\[
R \cdot K = {m_D^2 - m_{K^*\pi}^2 - m_{e\nu}^2\over 2}
\]
\[
\left|\vec k\right|^2 \equiv { (m_{K^*}^2 + m_\pi^2 - m_{K^*\pi}^2)^2 \over 4m_{K^*\pi}^2}
-{m_{K^*}^2 m_\pi^2 \over  m_{K^*\pi}^2}
\]
\[
\beta = {2\left|\vec k\right| \over  m_{K^*\pi}}
\]
\[
\chi = \sqrt{(R \cdot K)^2 - m_{K^*\pi}^2 m_{e\nu}^2}
\]
\[
K \cdot Q =  {m_{K^*}^2 - m_\pi^2 \over  m_{K^*\pi}^2} R \cdot K + \beta \chi \cos \theta_K
\]
\[
R \cdot L = \chi \cos \theta_e
\]
\[
Q \cdot L = \left({m_{K^*}^2 - m_\pi^2 \over  m_{K^*\pi}^2}\right)\chi \cos \theta_e
+ \beta (R \cdot K)\cos \theta_K \cos \theta_e - \beta m_{K^*\pi} m_{e\nu} \sin
\theta_K \sin \theta_e \cos \phi
\]
\[
2\eta \equiv p_D \cdot R = {m_D^2 + m_{K^*\pi}^2 - m_{e\nu}^2 \over 2}
\]
\begin{equation}
2\zeta \equiv p_D \cdot Q = (m_{K^*}^2 - m_\pi^2)\left( 1 + {R \cdot K \over m_{K^*\pi}^2}\right) + \beta \chi \cos \theta_K.
\label{dot2}
\end{equation}
In the above, $p_D$ denotes the four-momentum of the $D^0$.  
Also, in the limit where
$(m_\pi/m_{K^*}) \to 0$, we have $\chi \to X$, where $X$ is defined in Eq. (\ref{X}). 

We have the following expressions for the magnitude of the $\bar K^{*0}$
three momentum and the energy of the $\pi^-$, in the rest frame of the $D^0$,
denoted by $\left|\vec{p^\prime}\right|$ and $E_\pi$, respectively
\[
\left|\vec{p^\prime}\right| = \sqrt{\left({\eta + \zeta \over m_D}\right)^2  - m_{K^*}^2}
\]
\begin{equation}
E_\pi = {\eta - \zeta \over m_D}.
\label{Epi}
\end{equation}


\begin{thebibliography}{99}
\bibitem{Zoltan} This approach was brought to my attention by Zoltan Ligeti.
\end{thebibliography}
\end{document}